\documentclass[pra,twocolumn,a4paper,superscriptaddress,floatfix]{revtex4}

\usepackage{graphicx}
\usepackage{amsmath}
\usepackage{bm}
\usepackage{amssymb}
\usepackage{color}
\usepackage[normalem]{ulem}
\usepackage{dashrule}
\newcommand{\vect}[1]{{\mathbf #1}}

\newcommand{\blue}[1]{{\color{blue} #1}}
\newcommand{\green}[1]{{\color{green} #1}}
\newcommand{\ef}{\varepsilon_F}
\newcommand{\eb}{\varepsilon_B}
\newcommand{\wz}{\omega_z}
\newcommand{\ad}{a_{2\rm{D}}} 

\newcommand{\up}{\uparrow}
\newcommand{\down}{\downarrow}

\begin{document}

\title{Quasi-two-dimensional Fermi gases at finite temperature}

\author{Andrea M.\ Fischer}
\affiliation{Cavendish Laboratory, JJ Thomson Avenue, Cambridge, CB3 0HE,
United Kingdom}
\affiliation{London Centre for Nanotechnology, Gordon Street, London, WC1H 0AH,
United Kingdom}

\author{Meera M. Parish}
\affiliation{Cavendish Laboratory, JJ Thomson Avenue, Cambridge, CB3 0HE,
United Kingdom}
\affiliation{London Centre for Nanotechnology, Gordon Street, London, WC1H 0AH,
United Kingdom}
\date{\today}

\begin{abstract} 
We consider a Fermi gas with short-range attractive interactions that is confined along one direction by a tight harmonic potential. For this quasi-two-dimensional (quasi-2D) Fermi gas, we compute the pressure equation of state, radio frequency spectrum, and the superfluid critical temperature $T_c$ using a mean-field theory that accounts for all the energy levels of the harmonic confinement. 
Our calculation for $T_c$ provides a natural generalization of the Thouless criterion to the quasi-2D geometry, and it correctly reduces to the 3D expression derived from the local density approximation in the limit where the confinement frequency $\wz \to 0$. 
Furthermore, our results suggest that $T_c$ can be enhanced by relaxing the confinement and perturbing away from the 2D limit.
\end{abstract}


\maketitle
\section{Introduction}
\label{sec-introduction}
Two-dimensional (2D) Fermi systems are both of fundamental interest and
technological importance.
Classic examples include graphene \cite{CasGPN09}, high-temperature
superconductors \cite{TsuK00}, semiconductor interfaces~\cite{SmiM90}, and
layered organic superconductors \cite{SinM02}. 
In addition,
it is now possible to confine cold gases of alkali atoms in a 1D optical
lattice, leading to a series of quasi-2D layers \cite{ModFHR03,MarMT10}.
Here, the interlayer coupling can be tuned and indeed made
negligible by increasing the lattice depth \cite{SomCKB12}, thus allowing the investigation of single quasi-2D layers. 
Alternatively, interlayer tunnelling can be switched on and the behavior of simple layered systems investigated.  
Furthermore, the attractive short-range interactions between different fermionic species may be controlled using a magnetically tunable Feshbach resonance.
All these scenarios serve to illustrate the high degree of experimental control available in cold atoms, thus making them
ideal systems to study the behaviour of fermions in low dimensions.

Experiments on quasi-2D atomic Fermi gases have thus far focussed on the behavior of a single quasi-2D gas.
This is achieved by applying a sufficiently strong optical lattice so that 
interlayer tunnelling may be neglected, 
or by trapping the gas tightly in one direction 
(which we refer to as the $z$ direction).
In both cases,
the confining potential for the layer can be approximated by a harmonic
oscillator potential, i.e., $V(z)=\frac{1}{2}m\wz^2z^2$, where $m$ is the atom mass and $\wz$ is the confinement frequency~\footnote{There is also a considerably weaker harmonic confinement in the $x$,$y$ directions, with ${\omega_x=\omega_y \ll \wz}$, which we will ignore here. 
However,  one can map the properties of the uniform quasi-2D gas to the trapped gas using the local density approximation when the confinement is sufficiently weak.}.
The gas is considered to be kinematically 2D provided the Fermi energy $\ef$ and temperature $T$ satisfy
$\ef, k_B T \ll \hbar \wz$. In principle, the crossover from BCS pairing to tightly bound bosonic $\up\down$ dimers can then be realized in 2D by increasing the attractive $\up$-$\down$ interactions~\cite{randeria1989,RanDS90}, and this has been the subject of much investigation 
 \cite{FelFVK11,SomCKB12,ZhaOAT12,MakMT14}.
 Since a two-body bound state always exists in 2D for arbitrary attraction~\cite{LanL89}, the two-body binding energy, $\eb$, can be used to parameterize the interaction strength. In this manner, weak BCS pairing is achieved when $\eb/\ef\ll1$, while the Bose limit corresponds to 
 $\eb/\ef\gg 1$.
However,  in practice, it is difficult to remain strictly 2D when varying the parameter $\eb/\ef$. 
 In particular, many experiments in the BCS limit 
appear to be in the regime where $\ef \gtrsim \hbar \wz$~\cite{DykKWH11,ZhaOAT12,MakMT14} 
and therefore one expects measurable deviations from 2D behavior~\cite{MarT05,FisP13}.

In this work, we address the impact of confinement on a single, quasi-2D, two-component gas of fermions.
Previously, we developed a mean-field theory of the quasi-2D system at zero temperature that allowed us 
to extrapolate to 
an infinite number of harmonic confinement levels~\cite{FisP13}. Here, we extend our mean-field calculation 
to finite temperature and determine (i) the critical temperature for pair formation, $T_c$; (ii) the equation of state for the pressure; and 
(iii) the excitation spectrum obtained from radio frequency (RF) spectroscopy.
An earlier mean-field study of the regime $\ef \gtrsim \hbar \wz$ considered RF spectra at finite temperature, but not the equation of state or $T_c$ \cite{MarT05}. 
The pressure of the quasi-2D gas 
has recently been measured experimentally \cite{MakMT14}, and a comparison with finite-temperature \cite{BauPE13} and zero-temperature \cite{BerG11} calculations in 2D suggests that temperature noticeably increases the pressure in the BCS regime. 
We find here that confinement provides a competing effect, where an increase in $\ef/\hbar\wz$ can reduce the pressure.

To date, there is no experimental observation of the superfluid phase in atomic
2D Fermi gases. However, our results 
show that $k_B T_c/\ef$ is increased by relaxing the confinement and increasing 
$\ef/\hbar\wz$ for constant $\eb/\ef$, thus raising the tantalizing
possibility that
superfluidity is most favorable when the gas lies between 2D and 3D.
As far as we are aware, our work provides the first quasi-2D generalization of the Thouless criterion for $T_c$
as we perturb away from the 2D limit. 

\begin{figure}
  \centerline{
  \includegraphics[width=0.9\columnwidth]{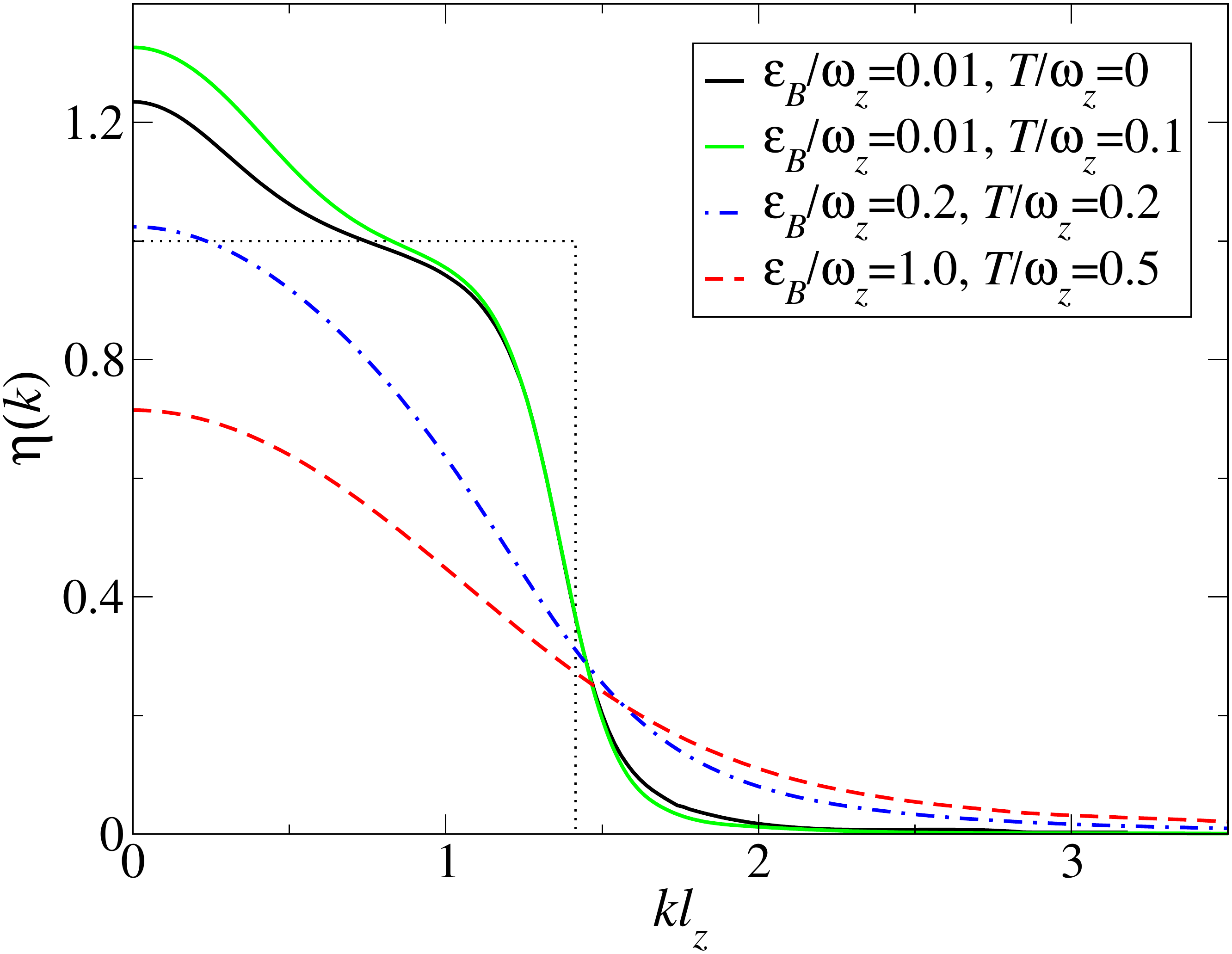} 
  }
  \caption{(Color online) The momentum distribution $\eta$ per
spin component, for Fermi energy $\ef=\wz$ and various binding energies
$\eb$ and
temperatures $T < T_c$. The momentum is scaled by the confinement length $l_z$. 
The black dotted line is the momentum
distribution for an ideal gas at zero temperature. 
 }
  \label{fig-momdist}
\end{figure}

\section{Model and Mean-field theory}
\label{sec-model}
We consider a balanced two-component ($\uparrow$, $\downarrow$) 
gas of fermions strongly confined in the $z$-direction by a harmonic
oscillator potential. 
Harmonic confinement within the
$x$-$y$ plane is not
included, so the atoms possess a 2D momentum, $\vect{k}$, in addition to an
oscillator quantum number, $n$. 
The many-body grand-canonical Hamiltonian in this case is
\begin{align}
\label{eq-ham}
\hat{H} 
= & \sum_{\mathbf{k}, n, \sigma}\xi_{\mathbf{k} n}
c^\dagger_{\mathbf{k} n \sigma} c_{\mathbf{k} n \sigma}\\
& +
\sum_{\substack{\mathbf{k}, n_1, n_2 \\ \mathbf{k}', n_3, n_4 \\ \mathbf{q}}}
\langle n_1 n_2 |\hat{g}| n_3 n_4 \rangle
c^\dagger_{\mathbf{k} n_1 \uparrow}c^\dagger_{\mathbf{q}-\mathbf{k} n_2
\downarrow}
c_{\mathbf{q}-\mathbf{k}' n_3 \downarrow}c_{\mathbf{k}' n_4 \uparrow}\nonumber,
\end{align}
where $\xi_{\mathbf{k} n}=\epsilon_{\mathbf{k} n}-\mu$, 
$\mu$ is the chemical potential,
and $\epsilon_{\mathbf{k} n}=k^2/2m +
n\omega_z$, corresponding to the single-particle energies relative to the zero-point energy 
$\omega_z/2$.
Here we set the system area ${\Omega =1}$ and ${\hbar=k_B=1}$. 
The short-range 3D interaction $\hat{g}$ is tuned by a broad $s$-wave Feshbach resonance. 
The interaction matrix elements are most easily calculated using
relative ($\nu$) and center of mass ($N$) oscillator quantum numbers:
\begin{align}
\label{eq-intmatelts}
 \langle n_1 n_2|\hat{g}| n_3 n_4 \rangle
&= g\sum_{N,\nu,\nu'} f_\nu\langle n_1 n_2|N \nu \rangle
f_{\nu'}\langle N \nu'|n_3 n_4\rangle  \nonumber \\
&\equiv g\sum_N V_N^{n_1 n_2}V_N^{n_3 n_4}.
\end{align}
Here, the function $f_\nu= (\frac{2\pi}{m\omega_z})^{1/4} \ \phi_\nu(0)$, with $\phi_\nu(z)$ being 
the $\nu$-th harmonic oscillator eigenfunction. One 
can show that ${f_{2\nu+1}=0}$ and
${f_{2\nu}=\frac{(-1)^\nu}{\nu!} \sqrt{\frac{
(2\nu)! } { 2^ { 2\nu } } } }$. The coefficients, $\langle n_1 n_2|N \nu
\rangle \sim \delta_{N+\nu \hspace{1mm} n_1+n_2}$, are determined in
Ref.~\cite{Smirnov1962,ChaW67}. The 
interaction strength 
$g$ can be expressed as a function of the binding energy $\eb$ of the
two-body problem:
\begin{align}
\label{eq-inverseg}
-\frac{1}{g} = \sum_{\textbf{k}, n_1,
n_2}\frac{(V_0^{n_1 n_2})^2}{\epsilon_{\textbf{k}
n_1}+\epsilon_{\textbf{k} n_2}+\eb} \>.
\end{align}
We can set $N=0$ without loss of generality, since the center of mass motion decouples. 
An equation relating the binding energy $\eb$ to the 3D scattering
length $a_s$~\cite{PetS01,BloDZ08} may be obtained using Eq.~(\ref{eq-inverseg}) together with the
renormalization condition
$\frac{1}{g}=\frac{m l_z}{\sqrt{2\pi}}\left(\frac{1}{2a_s}-\frac{\Lambda}{\pi} \right)$,
where $l_z = \sqrt{1/m\wz}$ and 
$\Lambda$ is a UV momentum cutoff in 3D.
Note that $g$ differs from the usual definition of the 3D interaction strength by a factor of $\sqrt{2\pi} l_z$.

Following the approach in Ref.~\cite{FisP13}, we define the superfluid order parameter
\begin{equation}
\label{eq-del}
\Delta_{\mathbf{q} N}=g\sum_{\mathbf{k}, n_1, n_2} V_N^{n_1 n_2} \langle
c_{\mathbf{q}-\mathbf{k}n_2\downarrow}c_{\mathbf{k}n_1\uparrow}\rangle.
\end{equation}
Provided fluctuations around this are small, we can approximate
Eq.~(\ref{eq-ham}) by the mean-field Hamiltonian
\begin{align}
\label{eq-mfham}
\hat{H}_\mathrm{MF} = & \sum_{\mathbf{k}, n, \sigma}\xi_{\mathbf{k} n}
c^\dagger_{\mathbf{k} n \sigma} c_{\mathbf{k} n \sigma}\\ \nonumber
& +
\sum_{\mathbf{q}, N}\bigg(
\Delta_{\mathbf{q} N}\sum_{\mathbf{k}, n_1, n_2}
V_N^{n_1 n_2}c^\dagger_{\mathbf{k} n_1 \uparrow}c^\dagger_{\mathbf{q}-\mathbf{k}
n_2
\downarrow}\\
& +
\Delta_{\mathbf{q} N}^\ast\sum_{\mathbf{k}, n_3, n_4}
V_N^{n_3 n_4}c_{\mathbf{q}-\mathbf{k}
n_3 \downarrow} c_{\mathbf{k}n_4 \uparrow}
-\frac{|\Delta_{\mathbf{q} N}|^2}{g}
\bigg).\nonumber
\end{align}
Furthermore, we assume that ${\Delta_{\mathbf{q} N}=\delta_{\mathbf{q}
\mathbf{0}}  \delta_{N 0}  \Delta_0}$. The assumption of a completely uniform order
parameter with $N=0$ should be reasonable in the limit where the pairing is weak or
$\Delta_0 \ll \ef$. 
It also correctly captures the short-distance behavior 
between pairs, thus allowing us to renormalize the 3D contact potential in a straightforward manner.
Equation~(\ref{eq-mfham}) can then be diagonalized to give
\begin{equation}
\label{eq-quasiham}
\hat{H}_\mathrm{MF}=\sum_{\mathbf{k}, n} (\xi_{\mathbf{k} n}-E_{\mathbf{k} n})
-\frac{\Delta_0^2}{g} +
\sum_{\mathbf{k}, n,
\sigma}E_{\textbf{k} n }\gamma^\dagger_{\textbf{k} n \sigma}
\gamma_{\textbf{k} n \sigma}, 
\end{equation}where 
$E_{\textbf{k} n }$ are the quasiparticle excitation energies. 
The quasiparticle creation and annihilation operators are
given by
\begin{align} 
\label{eq-gammaup}
\gamma^\dagger_{\textbf{k} n \uparrow} & =\sum_{n'} (u_{\textbf{k}
n'n}c^\dagger_{\mathbf{k} n' \uparrow}+
v_{\textbf{k} n'n}c_{-\mathbf{k} n' \downarrow}) \\ 
\label{eq-gammadown}
\gamma_{-\textbf{k} n \downarrow} & =\sum_{n'} (u_{\textbf{k} n' n
}c_{-\mathbf{k} n' \downarrow}-
v_{\textbf{k} n'n}c^\dagger_{\mathbf{k} n' \uparrow}) ,
\end{align}
where the amplitudes $u$, $v$ are assumed to be real, and they
satisfy ${\sum_{n'}
(|u_{\textbf{k}
n'n}|^2 + |v_{\textbf{k} n'n}|^2)=1}$. Both the quasiparticle energies and
amplitudes only depend on the momentum through its magnitude, $k
\equiv|\vect{k}|$. 

The BCS ground state corresponds to the quasiparticle vacuum, but at finite
temperature, quasiparticle states can be occupied and one must consider the 
mean-field grand potential: 
\begin{equation} 
\label{eq-haverage}
 \langle \hat{H}_\mathrm{MF} \rangle = \sum_{\mathbf{k}, n}
\left[
\xi_{\mathbf{k} n} 
-\frac{2}{\beta}\ln\left[
2 \cosh(\beta E_{\vect{k} n}/2)
\right]
\right]
-\frac{\Delta_0^2}{g}
\end{equation}
where $\beta\equiv T^{-1}$. 
For fixed $\mu$, the correct value of $\Delta_0$ is that which minimises $\langle
\hat{H}_\mathrm{MF} \rangle$. 

The Fermi energy in the quasi-2D system is defined to be the chemical potential 
of an ideal Fermi gas with the same particle density, i.e., one can write the particle density per spin as 
${\rho(\ef)=\frac{m}{2\pi}(n_h+1)(\ef-n_h\wz/2)}$,
where $n_h$ is the oscillator number of the highest occupied oscillator level for the non-interacting gas.
The density $\rho$ depends on the chemical potential through the quasiparticle
amplitudes according to
\begin{equation} 
\label{eq-density}
\rho \equiv \sum_\vect{k} \eta(\vect{k})=\sum_{\vect{k},n_1,n_2}\frac{e^{\beta
E_{\vect{k}
n_1}}|v_{\textbf{k} n_2n_1}|^2 +|u_{\textbf{k} n_2n_1}|^2}{1+e^{\beta
E_{\vect{k} n_1}}} 
\end{equation}
where $\eta(\vect{k})$ is the in-plane momentum distribution per spin component.

The presence of discrete energy levels in the quasi-2D confinement is clearly apparent in $\eta(\vect{k})$
when $\ef$ approaches $\wz$, as shown in Fig.~\ref{fig-momdist}. In particular, 
even when the gas is kinematically 2D at $T=0$, with a clear 2D distribution, we see that 
interactions can populate the higher levels, resulting in additional weight in the distribution around
${k=0}$. This feature is most pronounced when $\ef = \wz$, because here the Fermi energy touches the bottom
of the next band at $k=0$.
A finite temperature also leads to occupation of the higher levels, and it can thus 
enhance the peak at $k=0$.
However, large $T$ or large $\eb$ eventually smears out the distorted Fermi distribution.

\subsection*{Short-distance behavior} At large momentum values,
$\eta(\vect{k})$ is determined by the contact density $\mathcal{C}$ 
\cite{Tan08}, which gives a measure of the 
density of pairs at short distances. Similarly to 2D and 3D~\cite{Werner2012}, this can be derived from the adiabatic relation:
\begin{align}
\mathcal{C}  = \frac{m^2}{\hbar^4}
\frac{\partial \langle \hat{H}_\mathrm{MF} \rangle}{\partial
(-1/g)}\Bigr\rvert_{\Omega,\mu,T}
\end{align}
However, the tail of the momentum distribution in quasi-2D has a slightly modified behavior:
\begin{align}
\eta(\vect{k}) \to \frac{\mathcal{C}}{4m^2} \sum_\nu \frac{f_{2\nu}^2}{\epsilon_{\vect{k}\nu}^2}
\end{align} 
Note that this yields the expected 2D behavior when $\wz \gg k^2/2m$.
In the quasi-2D mean-field approximation, the contact is simply related to the pairing order parameter: 
$\mathcal{C}=m^2\Delta_0^2$. 
This expression captures the monotonic increase of $\mathcal{C}$ with increasing attraction, and it yields the correct two-body 
contact in the Bose limit. 
However, it is not quantitatively accurate in the BCS regime 
since the mean-field approach neglects interactions in the normal phase. 
Indeed, we see here that $\mathcal{C} = 0$ above $T_c$.

\section{Critical Temperature}
\label{sec-tc}

\begin{figure}
  \centerline{
  \includegraphics[width=0.95\columnwidth]{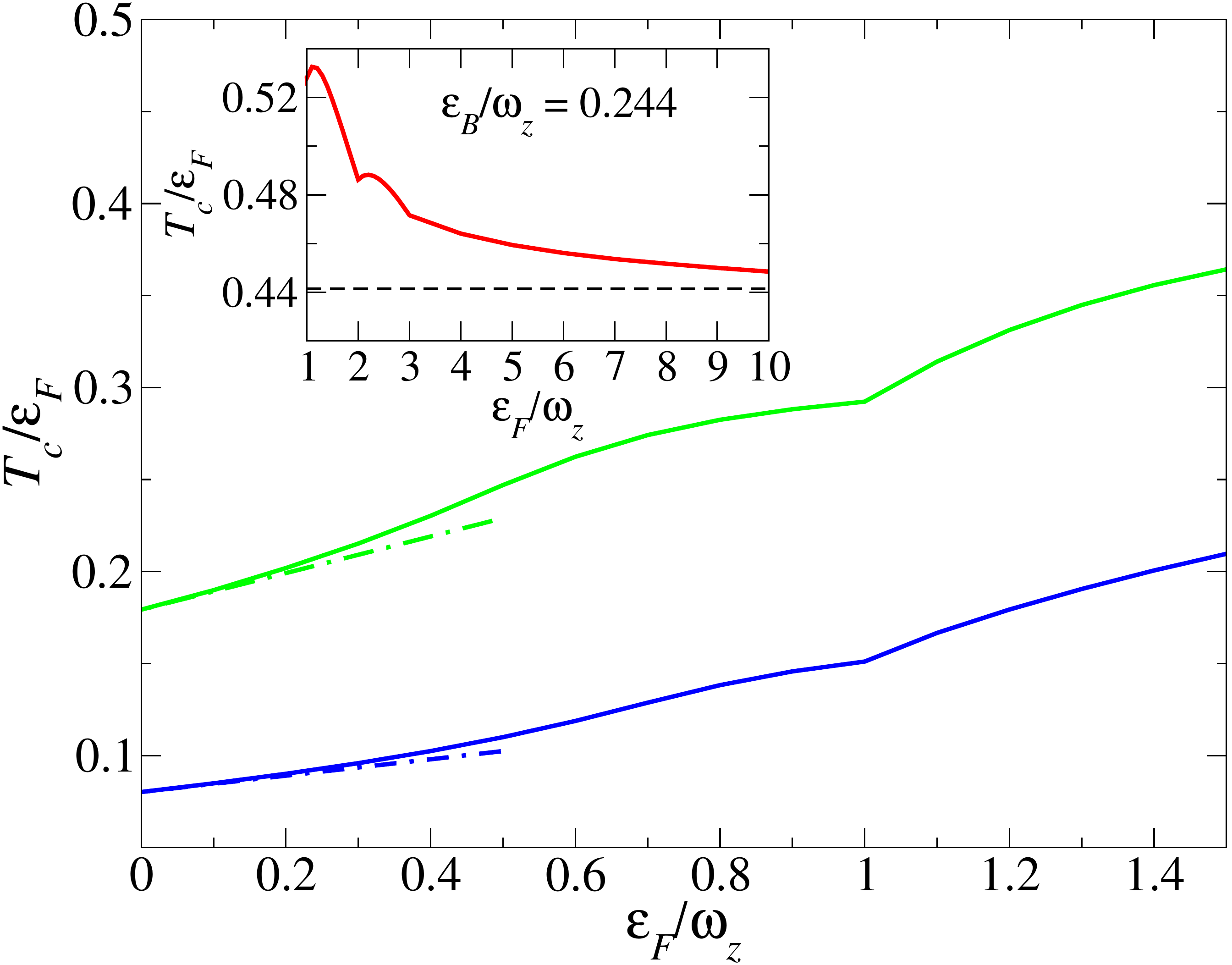} 
  }
\caption{(Color online)
Evolution of the critical temperature as the system is perturbed away from the 2D limit
for fixed values of the interaction, $\eb/\ef=0.01$
(\blue{\protect\rule[0.5ex]{3mm}{0.5mm}}) and $0.05$
(\green{\protect\rule[0.5ex]{3mm}{0.5mm}}).   
The dash-dotted lines are the leading order behavior in $\ef/\wz$ from Eq.~\eqref{eq-tcanalyt}.
Inset: The critical temperature at unitarity for large $\ef/\wz$.
The horizontal dashed line corresponds to the 3D limit $\ef/\wz \to \infty$ (see text).
}
  \label{fig-tc} 
\end{figure}

Within the mean-field approximation, pairing and superfluidity are destroyed simultaneously by thermal fluctuations, and 
the superfluid transition temperature corresponds to the point where $\Delta_0$ vanishes.
Here, we focus on the BCS regime of weak interactions, where our approximation is expected to be most accurate and 
the mean-field $T_c$ is close to the superfluid Berezinskii-Kosterlitz-Thouless (BKT) transition temperature~\cite{BotS06}.
First,  we recall how to derive an expression for $T_c$ in the 2D limit. The relevant equations are
(\ref{eq-inverseg}) and (\ref{eq-haverage}) with all sums over quantum number $n$ restricted
to the $n=0$ term. 
In this case, the quasiparticle energies are simply ${E_{\mathbf{k}
0}=\sqrt{\xi_{\mathbf{k}0}^2+\Delta_0^2}}$.
Substituting this into
Eq.~(\ref{eq-haverage}), and then 
using ${\partial \langle \hat{H}_\mathrm{MF} \rangle/\partial\Delta_0^2  = 0}$ and setting $\Delta_0=0$, 
yields the linearized gap
equation for $T_c$ (or Thouless criterion):
\begin{align}
{
\sum_{\textbf{k}}\frac{1}{2\epsilon_{\textbf{k}0}+\eb}
=\sum_\vect{k}\frac{
\mathrm{tanh}(\beta_c\xi_{\mathbf{
k
} 0 } /2) } { 2\xi_ {\mathbf{k}0}}
} \> . 
\end{align}
In the weak-coupling limit $\eb \ll \ef$, one can show that ${T_c=e^\gamma\sqrt{2\eb\ef}/\pi}$, where
$\gamma \approx 0.577$ is the Euler gamma constant \cite{miyake83}. The inclusion of 
Gor'kov--Melik-Barkhudarov corrections predicts a critical temperature
that is lower by a factor of $e$ \cite{PetBS03}.
 
The situation is more complicated in the general quasi-2D case. The gap equation is 
\begin{equation}
 \label{eq-gap}
-\frac{1}{g}=\sum_{\vect{k}, n}
\left. \left( 
\frac{\partial E_{\mathbf { k} n}}{\partial
\Delta_0^2}\right)\right|_{\Delta_0=0}
\tanh\left(\beta_c\xi_{ \mathbf { k} n } /2 \right)
\end{equation}
but there is no analytical expression for the quasiparticle energies. For
fixed $\vect{k}$, the $E_{\vect{k} n}$ are the positive eigenvalues
of the block matrix $M =
\left( \begin{smallmatrix} X&D\\ D^\dagger&-X \end{smallmatrix} \right)$, where
$X_{ij}=\delta_{ij}\xi_{\mathbf{k} i}$ and $D_{ij}=\Delta_0 V^{ij}_0$. Hence,
they are the solutions of the characteristic equation,
$|M- E_{\vect{k} n}I_n|=0$, where $I_n$ is the $n \times n$ identity matrix.
Note that the determinant contains only even powers of $\Delta_0$. Disregarding all
terms of order greater than two, we obtain
\begin{align}
\label{eq-det}
\left|M- E_{\vect{k} n}I_n\right| &
=
\prod_{n_1}\left(E_{\vect{k} n}^2-\xi_{\vect{k} n_1}^2\right)\nonumber \\
& -\Delta_0^2\sum_{n_1}
\left(V^{n_1n_1}_0\right)^2
\prod_{n_2\neq n_1}\left(E_{\vect{k} n}^2-\xi_{\vect{k} n_2}^2\right) \nonumber
\\
& + 2\Delta_0^2\sum_{n_1<n_2}\left(V^{n_1n_2}_0\right)^2
\left(\xi_{\vect{k} n_1}\xi_{\vect{k} n_2}-E_{\vect{k} n}^2\right)\nonumber \\
& \times \prod_{n_3\neq n_1,n_2}\left(E_{\vect{k} n}^2-\xi_{\vect{k}
n_3}^2\right).
\end{align}
After differentiating the expression $|M- E_{\vect{k} n}I_n|=0$ with respect to $\Delta_0$
and letting $\Delta_0 \to 0$, $E_{\vect{k} n}\to\xi_{\vect{k} n}$, we find
\begin{equation}
\left. \left(
\frac{\partial E_{\mathbf { k} n}}{\partial
\Delta_0^2}\right)\right|_{\Delta_0=0}
=\sum_{n'}\frac{\left(V^{nn'}_0\right)^2}{\xi_{\vect{k} n}+\xi_{\vect{k} n'}}.
\end{equation}
Hence, the linearized gap equation becomes
\begin{equation}
 \label{eq-gap2}
-\frac{1}{g}=\sum_{\vect{k}, n_1, n_2}
(V_0^{n_1 n_2})^2\frac{\mathrm{tanh}\left(\beta_c\xi_{ \mathbf { k} n_1 } /2
\right)+\mathrm{tanh}\left(\beta_c\xi_{ \mathbf { k} n_2 } /2
\right)}{2(\xi_{\vect{k} n_1}+\xi_{\vect{k} n_2})},
\end{equation}
which is a natural generalization of the corresponding equation in 2D. 
This is one of the main results of this paper.

We extract the behaviour in the limit ${\mu, \eb, T_c \ll \omega_z}$ 
by reorganising Eq.~\eqref{eq-gap2}:
\begin{multline}
\frac{1}{g} + \sum_{\vect{k},n_1,n_2} \frac{(V_0^{n_1 n_2})^2}{\xi_{\vect{k}n_1}
+ \xi_{\vect{k}n_2} +i0} = \\
\sum_{\vect{k},n_1,n_2} \frac{(V_0^{n_1 n_2})^2}{\xi_{\vect{k}n_1} +
\xi_{\vect{k}n_2} + i0} 
\left[ f(\beta_c\xi_{\vect{k}n_1}) + f(\beta_c\xi_{\vect{k}n_2}) \right],
\end{multline}
where $f(x)\equiv1/(1+e^{x})$ is the Fermi-Dirac distribution function.
We see that the left hand side is inversely proportional to the
quasi-2D $T$ matrix at energy $2\mu$, which can be readily expanded in
$1/\omega_z$~\cite{LevB12}.
Expanding both sides up to linear order in $1/\omega_z$ then yields
\begin{multline}
\label{eq-tclargewz}
\ln\left(\frac{\varepsilon_B}{2\mu}\right) + \ln(2) \frac{2\mu +
\varepsilon_B}{\omega_z}
\simeq \\
2\mathcal{P}\int_{-\beta_c\mu}^\infty d\epsilon \frac{f(\epsilon)}{\epsilon}+
\frac{4\ln\left[4(2-\sqrt{3})\right]}{\beta_c\omega_z} \ln(1+e^{\beta_c\mu}) ,
\end{multline}
where $\mathcal{P}$ denotes the principle value.
In the BCS weak-coupling 
limit $\eb\ll \ef$, we have $\mu\simeq \ef$ and $\beta_c \mu \to \infty$.
Thus, the integral term in
Eq.~(\ref{eq-tclargewz}) can be replaced by $-2\ln \left(2e^\gamma\beta_c\mu/\pi\right)$, 
and the whole equation can be simplified to give
\begin{equation}
 \label{eq-tcanalyt}
T_c=\frac{e^\gamma}{\pi}\sqrt{2\eb\ef}
\left[
1+ \frac{\ef}{\wz}\ln\left(\frac{7+4\sqrt{3}}{8}\right)
\right] .
\end{equation}
Therefore, for fixed interaction parameter $\eb/\ef$, we see that 
perturbing away from 2D actually \emph{increases} $T_c/\ef$.

To obtain results for larger values of $\ef/\wz$, we need to include multiple
oscillator levels and we must therefore proceed numerically.
Since the normal state is non-interacting in this approximation, 
we may determine the density (and $\ef/\wz$) for a given $\beta_c \mu$ and
$\beta_c \wz$ using the expression:
\begin{equation}
 \label{eq-num}
\beta_c \rho 
=\frac{1}{\pi}\sum_n\ln\left(e^{\beta_c(\mu-n \wz)}+1\right).
\end{equation}
Combining this with Eqs.~\eqref{eq-inverseg} and \eqref{eq-gap2}, we then 
solve for $\eb/\wz$ as a function of $\beta_c \wz$ and $\ef/\wz$.
The calculations are repeated for different numbers of levels, $\eb/\wz$ is
fitted as a function of the inverse number of levels, and the final value
assigned to $\eb/\wz$ is that obtained by extrapolating to an infinite number of levels.

The variation of $T_c/\ef$ with $\ef/\wz$ is shown in Fig.~\ref{fig-tc}. 
We clearly see that $T_c/\ef$ is enhanced by relaxing the confinement for fixed $\eb/\ef$, 
and the numerical results in the BCS regime are in excellent agreement with the analytical expression of Eq.~(\ref{eq-tcanalyt})
in the limit $\ef/\wz \ll 1$.
Moreover,  the enhancement of $T_c$ becomes even more pronounced with increasing $\ef/\wz$.
Interestingly, there are dips visible at low integer values of $\ef/\wz$. This is due to the
single particle density of states having a step-like structure and being
discontinuous at these points. 

Increasing $\ef/\wz$ further, we should eventually recover the result derived from 
treating the harmonic potential within the local density approximation (LDA), valid in the limit
$\ef/\wz \to \infty$. In this case, it is easy to show that Eq.~\eqref{eq-num} reduces to
\begin{equation}
 \label{eq-numlargeef}
\frac{m}{4\pi}\frac{\ef^2}{\wz}=\int
dz\sum_{\vect{k}_{\rm 3D}}\frac{1}{1+e^{\beta_c\left(k_{\rm 3D}^2/2m-\mu(z)\right)}} ,
\end{equation}
where $\vect{k}_{\rm 3D}$ is the momentum of the particles in 3D 
and ${\mu(z)=\mu-\frac{1}{2}m\wz^2z^2}$.
To determine the behavior of the gap equation in this limit, we first rewrite Eq.~(\ref{eq-gap2}) as
\begin{align} \notag
-\frac{1}{g}=\sum_{\vect{k}, \nu}
\frac{f_{2\nu}^2}{2 \xi_{ \mathbf{k} \nu }}
 \sum_{n_1} 
|\langle n_1 \hspace{1mm} 2\nu-n_1|0 \hspace{1mm} 2\nu\rangle|^2 
\tanh\left(\beta_c\xi_{ \mathbf { k} n_1 }/2
\right) . 
\end{align}
For $\wz\to 0$, we also require $\nu\to\infty$ to obtain a finite energy $\nu \wz$.
In this case, we can use Stirling's approximation for large values of $\nu$ to obtain
\begin{align}\notag
 \frac{1}{\sqrt{2\pi}l_z} \sum_{\nu=0}^{\infty} \frac{f_{2\nu}^2}{2 \xi_{ \mathbf{k} \nu }} \simeq 
\frac{1}{2\pi}\int_{-\infty}^\infty dk_z  \frac{1}{2 (\xi_{ \mathbf{k} 0} + k_z^2/2m)}
,
\end{align}
where we have  defined $\nu\wz = k_z^2/2m$.
We can also write ${|\langle n_1 \hspace{1mm} 2\nu-n_1|0 \hspace{1mm}
2\nu\rangle|^2=\frac{1}{2^{2\nu}} \begin{pmatrix} 2\nu \\ n_1
\end{pmatrix}
\approx \frac{1}{\sqrt{\pi\nu}}e^{-(n_1-\nu)^2/\nu}}$, for large $\nu$, 
so that 
$\sum_{n_1=0}^{2\nu}|\langle n_1 \hspace{1mm} 2\nu-n_1|0 \hspace{1mm}
2\nu\rangle|^2$ may be replaced by $\int_{-1/2}^{1/2}dx\delta(x)$, 
with $n_1 = 2\nu x + \nu$. 
Applying these approximations to Eq.~(\ref{eq-gap2})
and using the renormalization condition for $g$
finally gives
\begin{equation}
 \label{eq-unitarity}
\frac{m}{4\pi a_s} =
\sum_{\vect{k}_{\rm 3D}} \left[\frac{m}{k_{\rm 3D}^2}
-\frac{\tanh[\beta_c(k_{\rm 3D}^2/2m-\mu)/2]}{2(k_{\rm 3D}^2/2m-\mu)} \right]
\> ,
\end{equation}
which is exactly the LDA expression obtained by considering the Thouless criterion at the center
of the trap ($z=0$). The fact that Eq.~(\ref{eq-gap2}) reduces to the correct mean-field expression in the limit 
$\wz \to 0$ may be regarded as a validation of our approach.

Solving the LDA equations at unitarity ($1/a_s = 0$) gives 
the value $T_c/\ef=0.44$, as indicated in the inset of Fig.~\ref{fig-tc}. 
 As expected, our numerics converge to this value with increasing $\ef/\wz$.
Note that it is lower than the corresponding mean-field value in the 3D uniform case, $T_c^{\rm{3D}}/\ef=0.5$. 
%

\section{Pressure}
\label{sec-pressure}

\begin{figure}
  \centerline{
  \includegraphics[width=0.9\columnwidth]{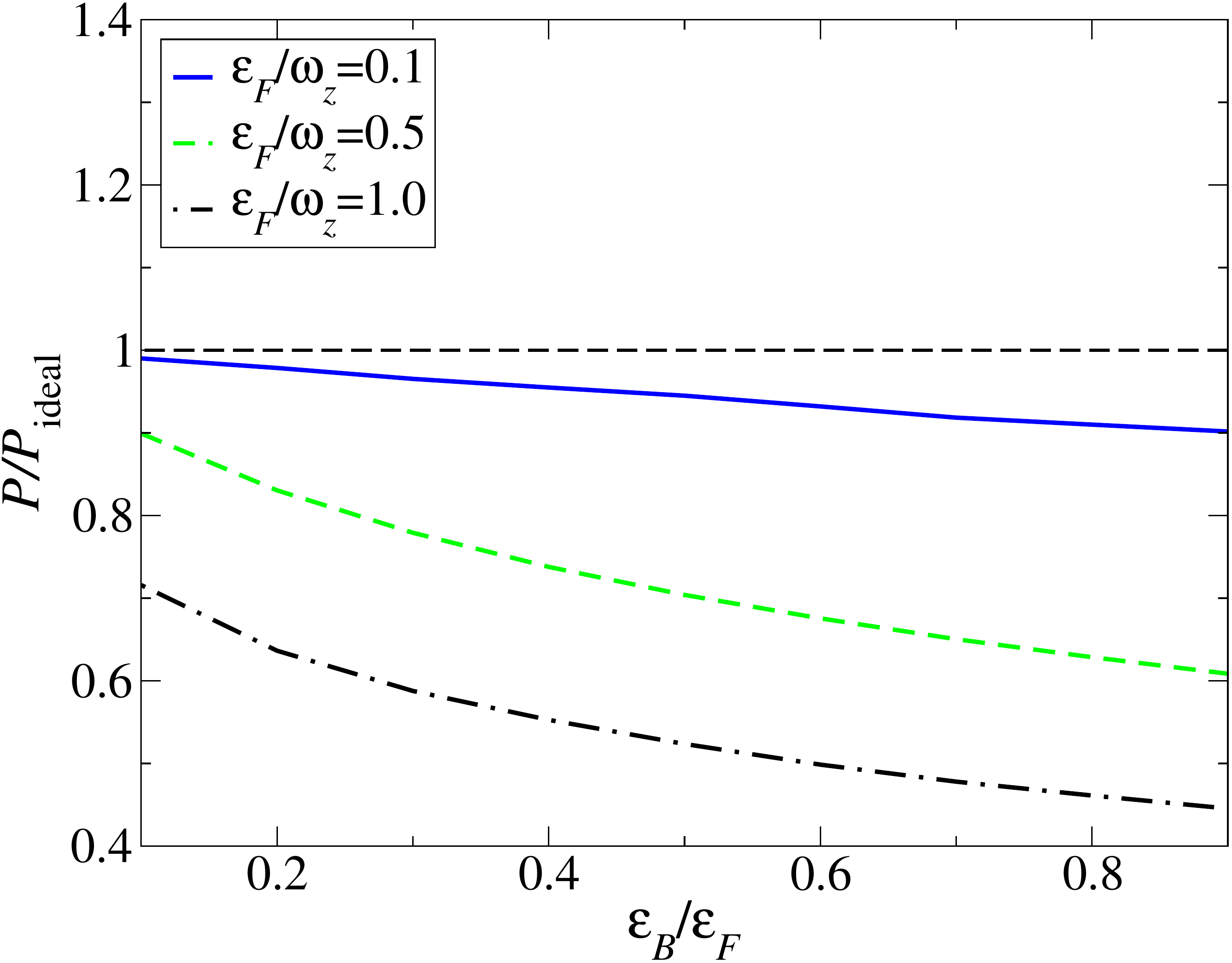} 
  }
 \caption{(Color online) Pressure of the quasi-2D gas 
for fixed Fermi energy $\ef/\wz$ and temperature $T/\ef =0.1$. 
Note that $T<T_c$ for the parameter range considered here.
The zero-temperature 2D mean-field pressure (dashed line) corresponds to that of an ideal 2D Fermi gas, $P_{\rm{ideal}}$.
The displayed interval of interactions corresponds to $\ad\sqrt{\rho} \approx 1$ (see text).}
  \label{fig-pressure}
\end{figure}

Below $T_c$, we can determine the mean-field equation of state for the pressure directly from ${P=-\langle
\hat{H}_\mathrm{MF} \rangle}$, once $\Delta_0$ is known. 
Like before, all calculations are repeated with different numbers of harmonic levels and 
the resulting pressure values extrapolated to an infinite number of levels.
In the 2D limit, the pressure at zero temperature corresponds to that of an ideal Fermi gas,
$P_{\rm{ideal}}=m\ef^2/2\pi$, and is therefore independent of interactions. 
This artifact is due to the fact that mean-field theory in 2D predicts an effective 
dimer-dimer repulsion that is classically scale invariant
(and thus only dependent on the density), rather than the correctly renormalized interaction 
expected for a quantum system~\cite{Pitaevskii1996,Olshanii2010}.
Once interactions beyond mean field are included, one finds that the pressure tends to zero 
with increasing attraction, $\eb/\ef$, since the system approaches 
a non-interacting Bose gas~\cite{BerG11,MakMT14}. 

Even though the mean-field approach is not accurate in pure 2D, 
we expect it to provide a reasonable estimate of the effect of confinement in the quasi-2D system, 
since mean-field theory becomes more reliable as we perturb away from 2D. 
Indeed, we see in Fig.~\ref{fig-pressure} that once $\wz$ is finite, the pressure in the plane decreases 
with increasing $\eb/\ef$, as expected. 
The interactions in quasi-2D may equivalently be parameterized by $\ad\sqrt{\rho}$, where
the scattering length  ${\ad=\sqrt{\pi/B}l_ze^{-\sqrt{\frac{\pi}{2}}\frac{l_z}{a}}}$
 and $B\approx 0.905$ \cite{PetS01}. 
 Increasing attraction and moving towards the Bose regime then corresponds to decreasing $\ad\sqrt{\rho}$. 
 In the 2D limit, we have $\ad = \sqrt{1/m\eb}$, but this is not true in general in quasi-2D. 
 The interaction range displayed in Fig.~\ref{fig-pressure} corresponds to the regime of 
 strong interactions $\ad\sqrt{\rho} \approx 1$ considered in a recent experiment~\cite{MakMT14}.
 Here, the pressure at finite temperature was observed to be lower than the 2D zero-temperature result.
However, this appears at odds with thermodynamics, where the pressure is always expected to increase with temperature.
Indeed, one can show for the 2D case that
\begin{align*}
\left( \frac{\partial P}{\partial T} \right)_{N,\Omega} 
= \left( \frac{\partial S}{\partial \Omega} \right)_{N,T} 
 \propto T \frac{\partial s}{\partial \tilde{T}} > 0 \> ,
\end{align*}
where $ \tilde{T} = T/\ef$ and $s=S/N$ is the entropy per particle. 
On the other hand, we see from Fig.~\ref{fig-pressure} that 
$P/P_{\rm{ideal}}$ becomes increasingly reduced as we relax the confinement, 
and this reduction can be substantial even when $\ef \leq \wz$.
Therefore, for the densities considered in the experiment, where $\ef \approx0.5\wz$~\cite{MakMT14}, 
we expect $P/P_{\rm{ideal}}$ at low temperatures to lie below the 2D zero-temperature result.
For weaker attraction, $P/P_{\rm{ideal}}$ is increased until eventually the effect of temperature dominates 
and we have $P/P_{\rm{ideal}}>1$.  
In this case, $P/P_{\rm{ideal}}$ will be higher than the  2D $T=0$ result, which is indeed what was observed~\cite{MakMT14}.

\section{Radio frequency spectra}
\label{sec-rfspec}

Radio frequency (RF) spectroscopy has been used in quasi-2D experiments to probe pairing 
and associated gaps in the energy spectrum~\cite{FroFVK11,SomCKB12,ZhaOAT12,BauFFV12}.
A pulse of RF laser light transfers atoms from one of the
initial hyperfine states (say $\left|\uparrow\right>$) to a third one that is previously unoccupied (denoted $|3\rangle$). The gas is then
released from the trap and the atoms in different hyperfine states separated,
so that the number of transferred atoms can be extracted. This is
repeated over a range of different frequencies to determine the RF spectrum. 

We calculate the transfer rate as a function of probe frequency using Fermi's golden rule. The perturbation to the Hamiltonian due to
the RF pulse is
\begin{equation}
 \label{eq-delh}
\delta \hat{H}\propto \sum_{\vect{k}, n}\left(
c^\dagger_{\mathbf{k} n 3}c_{\mathbf{k} n \uparrow}
+ c^\dagger_{\mathbf{k} n\uparrow}c_{\mathbf{k} n 3}
\right).
\end{equation}
Note that the initial and final state momenta are equal, owing to the long
wavelength of the RF radiation.
We also assume the ideal scenario where there are no final state interactions, 
which is reasonable for the case of $^{40}$K atoms~\cite{BauFFV12}.
In general, such interactions can modify the high-frequency behavior of the RF spectrum~\cite{LanBZB12}.

\begin{figure}
  
  \includegraphics[width=0.85\columnwidth]{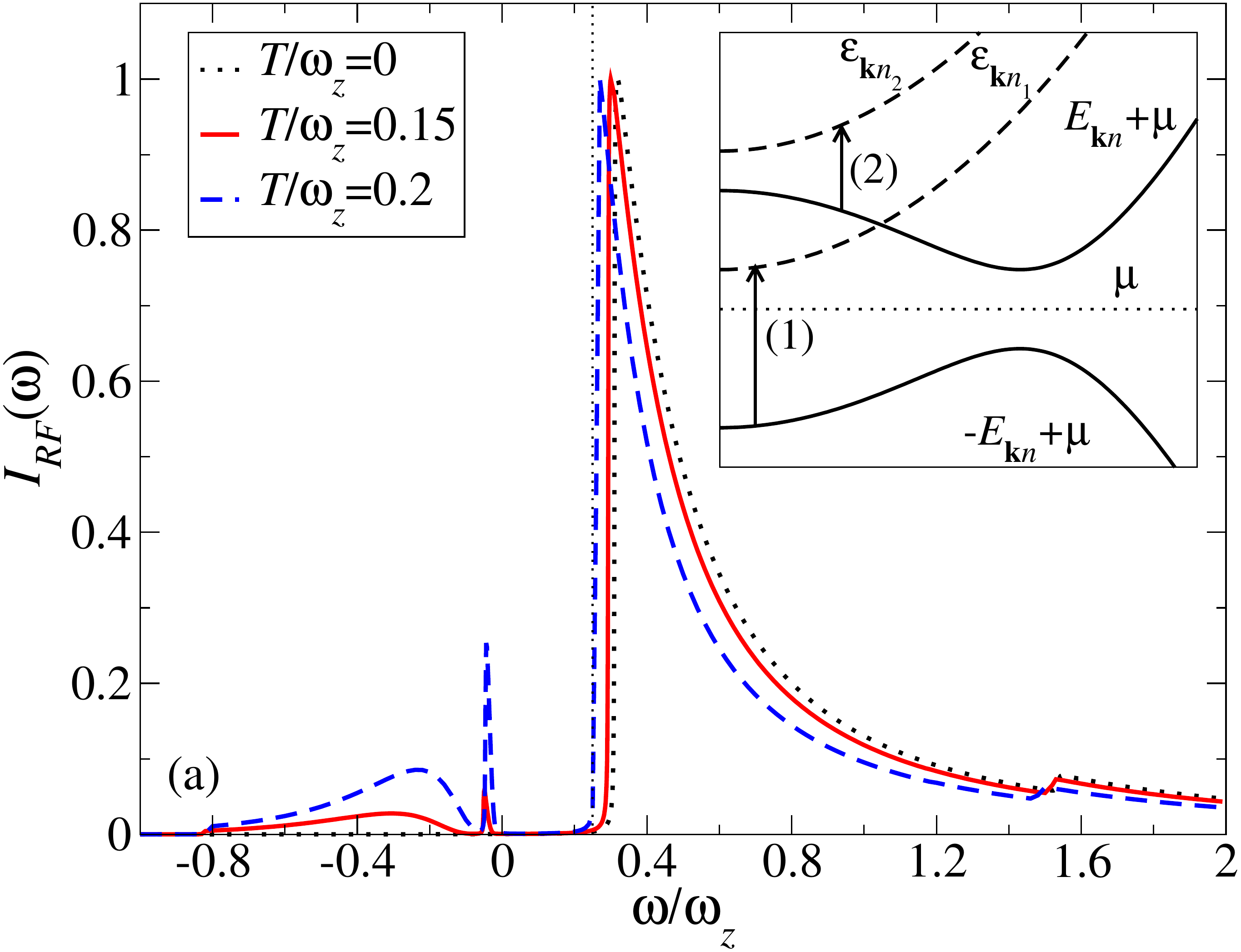}\\
  \includegraphics[width=0.85\columnwidth]{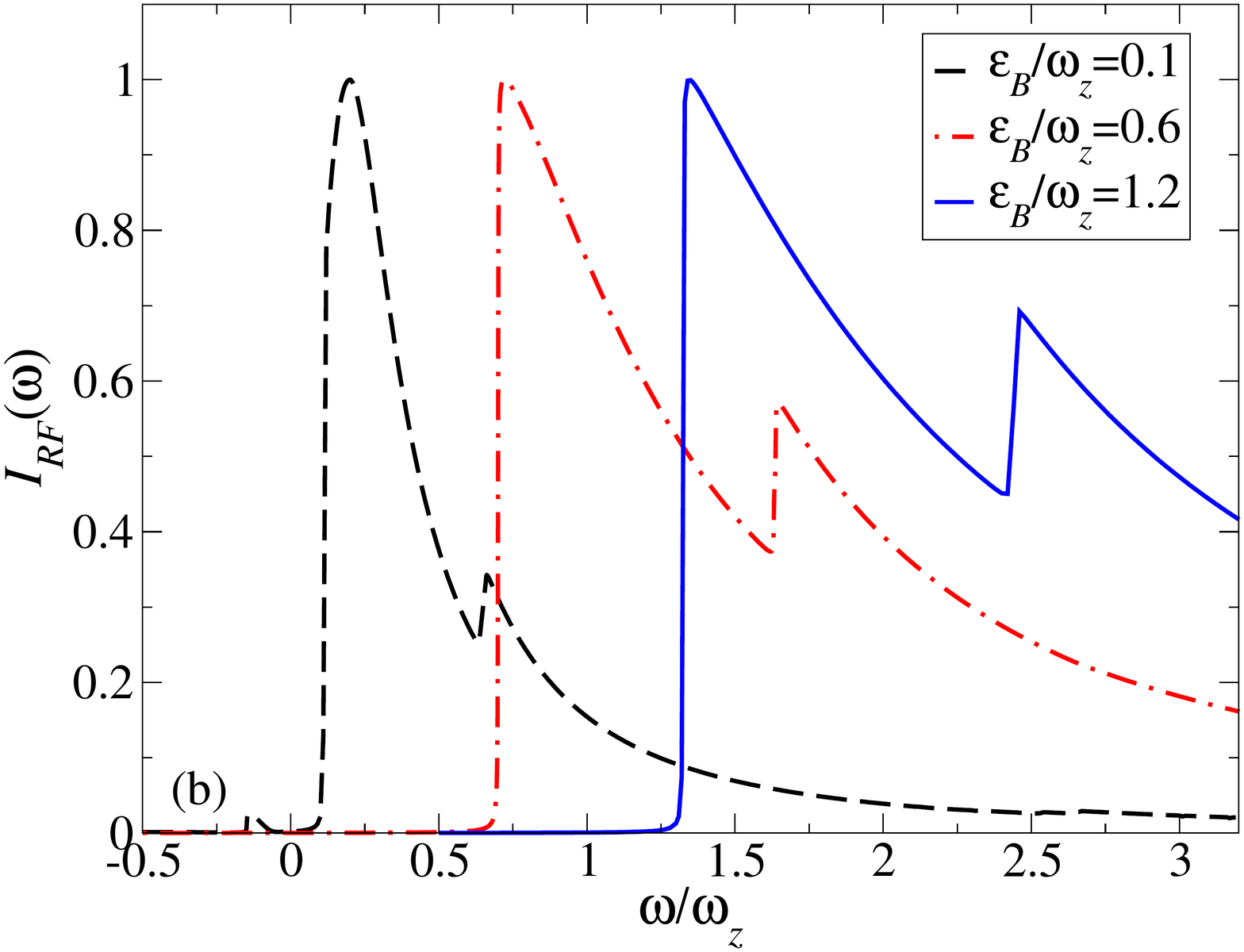} 
 
  \caption{(Color online) 
  RF spectroscopy plots for the quasi-2D Fermi gas at $T<T_c$.
  The RF current is scaled so that the largest peak is always has value 1.
  (a) RF spectra at various temperatures for $\eb/\wz=0.25$, $\ef/\wz=0.5$. 
  The vertical dotted line marks $\omega = \eb$.
Inset: Different types of ``bound'' [solid line] to free [dashed line] transitions (see text). 
(b) RF spectra at low temperature $T/\ef=0.1$ for $\ef/\wz=1.0$ and a range of binding energies.}
  \label{fig-rfa}
\end{figure}

There are two possible types of transition, as illustrated in the inset of Fig.~\ref{fig-rfa}(a).
At zero temperature, the initial state is the BCS groundstate, 
$|\Psi_0\rangle \propto \prod_{\vect{k}n\sigma}
\gamma_{\vect{k}n\sigma} |0\rangle$, where $|0\rangle$ is the vacuum for the
operators $c_{\vect{k}n\sigma}$. One may think of $|\Psi_0\rangle$ as a filled sea of
``quasiholes'', each having energy $-E_{\vect{k} n}+\mu$, with
$\gamma_{\vect{k}n\sigma}$ the creation operator for a quasihole. Transition (1) corresponds to a quasihole being
destroyed and an atom created in a free state. Hence the final state is given
by:
$|\Psi_F^{(1)}(\vect{k},n_1,n_2)\rangle = c^\dagger_{
\mathbf{k} n_1 3}
\gamma^\dagger_{-\vect{k}n_2 \downarrow}|\Psi_I\rangle$.
At zero temperature, only transitions of type (1) are possible. 
However, at finite temperature, the initial state $|\Psi_I\rangle$ also contains some
quasiparticles. Therefore, we can also have transition (2), where
a quasiparticle is destroyed and an atom is created in a free
state. The final state in this case is given by:
$|\Psi_F^{(2)}(\vect{k},n_1,n_2)\rangle = c^\dagger_{\mathbf
{k} n_1 3}
\gamma_{\vect{k}n_2 \uparrow}|\Psi_I\rangle$.
Accounting for both of these bound-to-free transitions, the mean-field RF current or transition rate is then
\begin{align} \notag
 \label{IRF}
I_{\rm{RF}}(\omega) &
\propto  \sum_{\substack{\vect{k}\\ n_1,n_2}}
\left[
f(-\beta E_{\vect{k}n_2}) \delta(\xi_{\mathbf{k} n_1}+E_{\vect{k}n_2}-\omega)
|v_{\vect{k} n_1 n_2}|^2 \right.  \\ 
& \left.+
f(\beta E_{\vect{k}n_2}) \delta(\xi_{\mathbf{k} n_1}-E_{\vect{k}n_2}-\omega)
|u_{\vect{k} n_1 n_2}|^2 \right] .
\end{align}
Here, the pulse
frequency relative to the bare transition frequency between hyperfine states is
denoted by $\omega$. 

We can gain insight into the RF spectrum by restricting the calculation to the two lowest harmonic oscillator levels,
$n=0,1$. This two-level approximation provides a qualitative picture of the quasi-2D spectrum at finite temperature and
has the advantage of yielding an analytical expression: 
\begin{align}
 \label{eq-rf2level}
 I_{\rm{RF}}(\omega) & \propto  \sum_{n=0,1}
\frac{(\Delta_0V_0^{nn})^2}{\omega^2}f\left(-\beta\frac{\omega^2+(\Delta_0
V_0^{nn})^2}{2\omega} \right) \\
& \times
\bigg[\Theta\left(\omega-\xi_n-\sqrt{\xi_n^2+(\Delta_0 V_0^{nn})^2}\right)
\nonumber \\
&
+ \Theta\left(-\omega\right)\Theta\left(\omega-\xi_n+\sqrt{\xi_n^2+(\Delta_0
V_0^ {nn})^2 } \right)\bigg] , \nonumber
\end{align}
where $\xi_n = n\wz -\mu$. We see immediately that $ I_{\rm{RF}}$ scales with $\Delta_0^2$,
and that there are contributions to the spectrum at both positive and negative frequencies
once temperature is finite.
Note there is no transition between the $n=0$ and $n=1$
levels, since parity must be conserved.

Figure \ref{fig-rfa} displays the numerical result for the RF spectra involving multiple harmonic levels. 
We focus on the regime $\ef \leq \wz$ where the spectrum has a simpler structure~\cite{FisP13}.
We have checked that the results have converged by varying the number of harmonic oscillator levels 
in the calculation of $u_{\vect{k} n_1 n_2}$ and $v_{\vect{k} n_1 n_2}$. 
For the frequency range displayed here,
we see that there are always two peaks at positive values of $\omega$. The
dominant peak is due to transitions of type (1) within the $n=0$ oscillator
level. The first term in Eq.~(\ref{eq-rf2level}) predicts a sharp onset of this peak at
${\omega=(\epsilon_{\vect{k}0}+E_{\vect{k}0}-\mu)|_{k=0}=\eb}$. However, the
numerical results show a confinement-induced shift to higher frequencies,
consistent with what we found previously~\cite{FisP13}.
The second peak at higher $\omega$ is due to transitions of type (1) within the
$n=1$ oscillator level. As expected, it is much weaker when $\wz$ is the largest energy scale.
However, this secondary peak is enhanced for larger $\ef$ and grows with increasing $\eb$, as shown in Fig.~\ref{fig-rfa}(b). 
 At finite temperatures, 
 two peaks also appear at negative $\omega$ values, due to type (2) transitions (see second
term in Eq.~(\ref{eq-rf2level})). In Fig.~\ref{fig-rfa}(a), the more rounded peak at lower frequency is
from transitions in the $n=0$ level and the sharper peak close to $\omega=0$ is
from transitions in the $n=1$ level. 

We can investigate the behavior at large $\eb/\ef$ by considering the two-body problem in quasi-2D. 
Using the dimer wave function $\phi_\vect{k}^{n_1 n_2} \propto  V_0^{n_1 n_2}/(\epsilon_{\vect{k} n_1}+\epsilon_{\vect{k} n_2}+\varepsilon_B)$ with zero center-of-mass momentum in the plane,
we obtain the simple expression
\begin{align}
 I_{\rm{RF}}(\omega) & \propto \frac{1}{\omega^2} \sum_\nu f^2_{2\nu} \Theta\left(\omega -\eb -2\nu\wz \right)
\end{align}
Thus, we see that we have peaks at $\omega = \eb +2\nu\wz$, similar to the structure 
displayed in Fig.~\ref{fig-rfa}(b). The secondary peaks grow in size with increasing $\eb/\wz$ until eventually
one recovers $ I_{\rm{RF}}(\omega) \propto \Theta(\omega -\eb) \sqrt{\omega - \eb}/\omega^2$, 
the behavior expected in 3D~\cite{Haussmann2009}.

\section{Conclusion}
\label{sec-conc}

We have considered the effect of a quasi-2D geometry on a Fermi gas at finite temperature. 
Such a study is important for ongoing experiments aimed at investigating the 2D BCS-Bose crossover, 
since many are in the regime where the confinement frequency $\wz$ is a relevant energy scale. 
In particular, the pressure can be substantially reduced by increasing $\ef/\wz$, and this appears to be
consistent with recent measurements~\cite{MakMT14}.
Furthermore, once $\ef \gtrsim \wz$ or $\eb \gtrsim \wz$, the discrete nature of the quasi-2D confinement should be clearly visible 
in the pairing properties even at finite temperature, as we can see from Figs.~\ref{fig-tc} and \ref{fig-rfa}.
A recent experiment at temperatures well above $T_c$  
has already observed steps in the cloud aspect ratio for integer $\ef/\wz$~\cite{DykKWH11}.

Rather than being an unwanted complication, the effects of confinement may provide a route to realizing 
superfluidity in quasi-2D Fermi gases. We have shown here that increasing $\ef/\wz$ for fixed interaction $\eb/\ef$ can actually increase 
$T_c/\ef$; thus it remains to be seen whether $T_c$ can be maximized with a geometry that lies between 2D and 3D. 
This requires a calculation that goes beyond mean-field theory and includes center-of-mass fluctuations of the pairs, since the size of $T_c$ eventually becomes too large to 
be determined only by pair breaking excitations.

In general, one will require approaches beyond the mean-field approximation in order to fully characterize the quasi-2D system.
Normal-state interactions are an obvious omission in the theory and are expected to impact the RF spectra~\cite{Pie12}.
There is also the possibility of a pseudogap regime just above $T_c$, where both bosonic pairs and Fermi statistics are present
\cite{FelFVK11,NgaLP13,BauPE13}. 
Nonetheless, our mean-field theory provides a major step towards including the effects of confinement.

\begin{acknowledgments}
We gratefully acknowledge fruitful discussions with Selim Jochim, Michael K\"ohl, Jesper Levinsen, and Stefan Baur.
This work was supported by the EPSRC under Grant No.\ EP/H00369X/2.
\end{acknowledgments}

\bibliographystyle{prsty}
\bibliography{bibliograph}
\end{document}